\documentclass[twocolumn,superscriptaddress,notitlepage,prb,showpacs,preprintnumbers,amsmath,amssymb]{revtex4}
\usepackage{amsmath}
\usepackage{amssymb}
\usepackage{dcolumn}
\usepackage{graphicx}
\usepackage{bm}

\begin{document}
\title{A Monte Carlo Approach for Studying Microphases Applied to the Axial Next-Nearest-Neighbor Ising and the Ising-Coulomb Models}
\author{Kai Zhang} 
\affiliation{Department of Chemistry, Duke University, Durham, North
Carolina, 27708, USA}
\author{Patrick Charbonneau}
\affiliation{Department of Chemistry, Duke University, Durham, North
Carolina, 27708, USA}
\affiliation{Department of Physics, Duke University, Durham, North
Carolina, 27708, USA}
\date{\today}

\begin{abstract}
The equilibrium phase behavior of microphase-forming systems is notoriously difficult to obtain because of the extended
metastability of their modulated phases. In this paper we present a systematic simulation methodology for studying layered microphases and apply the approach to two prototypical lattice-based systems: the three-dimensional axial next-nearest-neighbor Ising (ANNNI) and Ising-Coulomb (IC) models. The method involves thermodynamically integrating along a reversible path established between a reference system of free spins under an ordering field and the system of interest. The resulting free energy calculations unambiguously locate the phase boundaries. The simple phases are not observed to play a particularly significant role in the devil's flowers. With the help of generalized order parameters, the paramagnetic-modulated critical transition of the ANNNI model is also studied. We confirm the XY universality of the paramagnetic-modulated transition and its isotropic nature. Interfacial roughening is found to play at most a small role in the ANNNI layered regime.

\end{abstract}
 \maketitle



\section{introduction}
Lattice models are central to statistical mechanics. They strip
away the complexity due to packing and help reveal the influence of non-geometrical factors on both equilibrium and non-equilibrium self assembly.
The Ising model, for instance, offers a singular window on critical phenomena and on gas-liquid
coexistence~\cite{chandler:1987}; Flory-Huggins's theory of solvated polymers is core to the
physics of polymers~\cite{degennes:1979}; and spin glasses are key sources of inspiration for the difficult problem of structural glass formation~\cite{parisi:2010}. If a lattice model of a system exists, it is often a good strategy to solve it before embarking on a study of more elaborate variants.

Microphase formation is one such phenomenon that could benefit from further consideration of lattice-based models.
The frustration of short-range attraction -- or sometimes repulsion~\cite{glaser:2007} -- by a long-range
repulsion, irrespective of the physical and chemical nature of these interactions, leads to
universal spatially modulated patterns~\cite{seul:1995}. Periodic lamellae, cylinders, clusters, etc.~are thus similarly found in block
copolymers~\cite{hamley:1998,leibler:1980,jenekhe:1999}, oil-water surfactant mixtures~\cite{wu:1992,gompper:1992}, charged colloidal suspensions~\cite{stradner:2003}, and numerous magnetic materials~\cite{rossat-mignod:1980,seul:1992}. Microphase formation has also
been hypothesized to play a role in biological membrane organization~\cite{sieber:2007} and in the formation of
stripes in certain superconductors~\cite{tranquada:1995,emery:1999,vojta:1999,orenstein:2000,vojta:2009,parker:2010}, though the microscopic interpretation is still
debated. The spontaneous nature of microphase organization allows for these mesoscale periodic textures to find technological success as thermoplastic elastomers~\cite{hamley:1998} and nanostructure templates~\cite{thurn:2000}. Obtaining a detailed control over microphase morphology remains, however, notoriously difficult~\cite{cohen:1990}.
Annealing~\cite{leung:1986}, external fields~\cite{koppi:1993}, strain compression~\cite{kofinas:1994}, addition of fullerenes~\cite{jenekhe:1998,jenekhe:1999}, or
complex chemical environments~\cite{meli:2009} are often necessary to order diblock copolymers, for instance.

Understanding how to tune and stabilize microphases is essential to broadening their material relevance, yet experimental systems provide limited microscopic
insights. A number of continuous space~\cite{sear:1999,pini:2000,cao:2006,archer:2007,klix:2009,klix:2010,bomont:2010} and
lattice~\cite{kretschmer:1979,selke:1988,yeomans:1988,selke:1992,widom:1986,fried:1991,matsen:2006,low:1994,tarjus:2001} models have thus been devised for
theoretical and simulation studies, and some of which have even become textbook material~\cite{chaikin:1995,landau:2000}. Grasping the equilibrium properties of these models is necessary
to resolve problems surrounding the non-equilibrium assembly of microphases~\cite{cates:1989,charbonneau:2007,toledano:2009}.
But although the modulated regime is a key feature of these models, it has not been accurately characterized in any of them. Even for the most schematic formulations,
the existing theoretical treatments have only offered limited assistance.

Direct computer simulations have also been unable to provide  reliable equilibrium information~\cite{micka:1995,landau:2000}.
Traditional simulation methodologies that facilitate ergodic sampling of phase space by passing over free energy
barriers, notably parallel tempering and cluster moves, are of limited help in microphase-forming
systems. Because of the dependence of the equilibrium periodicity on temperature, sampling higher temperatures
leaves the system in a modulated phase with the wrong periodicity; and because of the lack of simple structural
rearrangements for sampling different modulations, the efficiency of cluster moves is limited. We recently
introduced a free-energy integration method for simulating modulated phases that overcomes this
hurdle~\cite{zhang:2010}. Here, we detail this method and apply it to the study of two canonical
three-dimensional (3D) spin-based systems: the axial next-nearest-neighbor Ising (ANNNI) and the Ising-Coulomb
(IC) models. Both of these models are known to form lamellar phases of different periodicities at low
temperature, but their phase structure is still not completely understood. The phase information we obtain by
simulation further allows testing of various theoretical predictions. The plan of this paper is to introduce the models
(Sect.~\ref{sect:models}), the simulation methodology (Sect.~\ref{sect:method}), and the generalized order and critical parameters (Sect.~\ref{sect:order}). After discussing the results (Sect.~\ref{sect:results}), a short conclusion follows.

\section{Models}
\label{sect:models}
\begin{figure}
\includegraphics[width=2in]{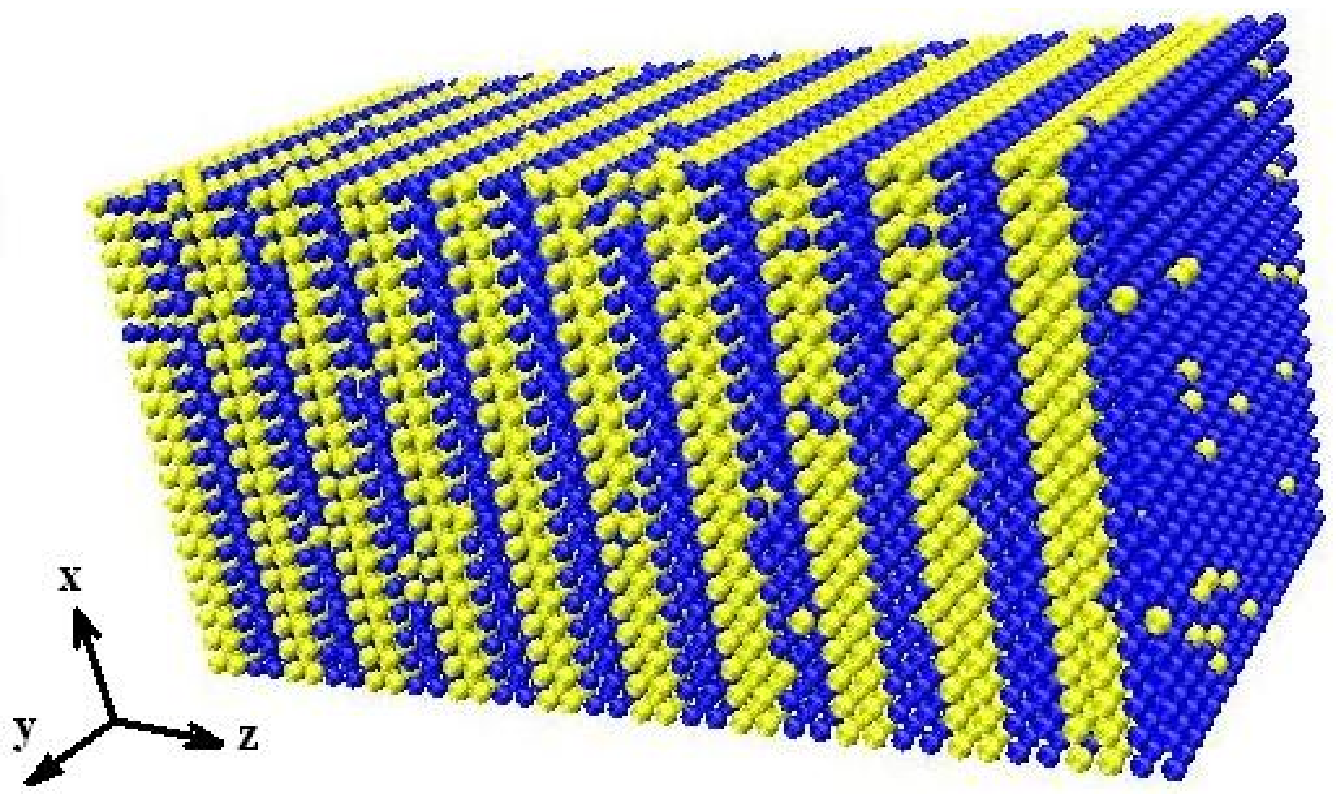}
\includegraphics[width=3.5in]{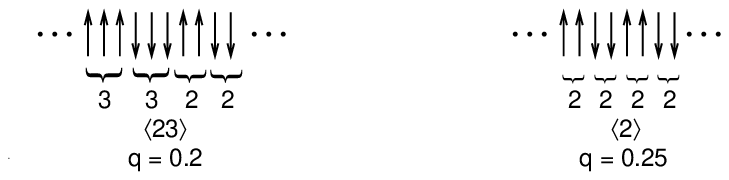}
\caption{(Top) Snapshot of the ANNNI antiphase $\langle2\rangle$ at $T = 2.4$, $\kappa = 0.7$ for a $20\times20\times40 $ lattice. Differently shaded beads indicate spins up or down.
(Bottom) Notation examples of the lamellar phases $\langle23\rangle$ and $\langle2\rangle$.} \label{fig:notation}
\end{figure}
Before introducing the models, a clarification of the nomenclature for describing layered microphases is in order.
Two conventions for characterizing the periodicity of lamellar phases coexist in the scientific literature.
The first compactly identifies a phase with a simple wave number $q=1/\lambda$ (in units of $2\pi$), where $\lambda$ is
the period length. The second, a short-hand form $\langle m^jn^k\rangle$ introduced in Ref.~\onlinecite{selke:1979}, is less compact but provides a more intuitive description of the layered phase.  In this notation, integers are used to describe a lamellar phase formed by periodic repetition of patterns
of $j$ lamellae of width $m$ followed by $k$ lamellae of width $n$ (Fig.~\ref{fig:notation}). For example, phase $\langle\infty\rangle$ is the ferromagnetic phase, phase $\langle2\rangle$ consists of two layers
of spins up followed by two layers of spins down, and phase $\langle 23\rangle$ has a period of 5.
Because thermal fluctuations blur the layer boundaries, the thickness of each lamella is generally not an integer but takes an average value
\begin{equation}
\frac{\lambda}{2}=\frac{mj+nk}{j+k}.
\end{equation}
This notation, which can only represent phases of rational periodicity, is well suited for the commensurate phases that are here observed.

\subsection{ANNNI Model}
The ANNNI model was first introduced to rationalize
helical magnetic order in certain heavy rare-earth
metals~\cite{elliott:1961,selke:1988,yeomans:1988,selke:1992}. The simple model's description of the experimentally observed order is only qualitative~\cite{muraoka:2002}, but because of its surprisingly complex phase behavior, it is now canonical for the study of systems with competing interactions~\cite{chaikin:1995,landau:2000}. Its Hamiltonian on a simple cubic lattice
\begin{equation}
H_{\mathrm{ANNNI}} =-J\sum_{\langle i,j\rangle}s_{i}s_{j}+\kappa
J\sum_{[i,j]_z}s_{i}s_{j}
\end{equation}
is expressed for spin variables $s_i=\pm 1$ coupled through a positive constant $J$. With the Boltzmann constant $k_B$, $J/k_B$ sets the temperature
$T$ scale. Alignment is favored for nearest-neighbor pairs $\langle i,j\rangle$, but frustrated with relative strength
$\kappa>0$ for $z$-axial next-nearest-neighbor pairs $[i,j]_z$. The exact solution of the one-dimensional version of the model provides $T=0$ phase information for all other dimensions~\cite{selke:1979}: ferromagnetic order is
the ground state for $\kappa<1/2$, while the layered antiphase $\langle 2 \rangle$ minimizes the energy for
$\kappa>1/2$. A mean-field description qualitatively captures the higher-dimensional, finite-$T$ features of the
model~\cite{jensen:1983,selke:1984}: the system is paramagnetic at high $T$; it is ferromagnetic at low $T$
and $\kappa$; and modulated layered phases form for sufficiently high $\kappa$~\cite{selke:1992}. These three
regimes join together at a multicritical Lifshitz point $(\kappa_L, T_L)$ whose
special critical properties have been predicted by theory~\cite{hornreich:1975,diehl:2000,diehl:2002} and verified in
simulations~\cite{pleimling:2001,henkel:2002}. High-temperature series expansions have also been used to study the
paramagnetic phase and predict its limit of stability~\cite{stanley:1977a,oitmaa:1985}. These predictions were
confirmed by finite-size critical rescaling for the paramagnetic-ferromagnetic (PF)
transition~\cite{selke:1979,selke:1980,rasmussen:1981,kaski:1985}  and by heat capacity~\cite{selke:1979,rotthaus:1993} and generalized susceptibility~\cite{zhang:2010} measurements for the paramagnetic-modulated (PM)
transition. For $\kappa<\kappa_L$, the PF
transition has Ising universality~\cite{selke:1978,kaski:1985}; while for $\kappa>\kappa_L$, the PM transition has been argued to have XY
universality~\cite{garel:1976,droz:1976,selke:1988}, but direct simulation verifications are
incomplete~\cite{zhang:2010} and the results of the high-temperature series expansion analysis are
inconclusive~\cite{oitmaa:1985,mo:1991}. The ferromagnetic-modulated (FM) transition is predicted by a Landau-Ginzburg treatment
to be first order with $q$ changing discontinuously from $0$, and to be tangent
to the PF and PM transition lines at the Lifshitz point~\cite{michelson:1977a}.

A sequence of commensurate $\langle 2^j3\rangle$ phases spring from the multiphase point at $T=0$ and $\kappa=1/2$. The structure of the branching processes at low $T$ has been carefully studied~\cite{fisher:1980}, and forms the basis for the low-temperature series expansion~\cite{fisher:1981}. For the rest of the modulated regime, approximate theoretical treatments, such as an approximate mean-field theory with a soliton correction~\cite{bak:1980}, an effective-field theory~\cite{surda:2004}, and the tensor product variational approach~(TPVA)~\cite{gendiar:2005}
have been used. Monte Carlo simulations have also been carried out in this regime~\cite{selke:1979,rasmussen:1981},
but the hysteresis resulting from the high free-energy barriers that separate modulated phases from each other
limits accurate determinations of the phase boundaries from annealing-based
approaches~\cite{yeomans:1988,zhang:2010}. Avoiding annealing is thus preferable for accurately locating transitions within the modulated regime~\cite{zhang:2010}. It is thought that incommensurate
phases could lower the transition free energy barriers between different commensurate modulated phases on sufficiently
large lattices~\cite{fisher:1980}, but these phases have not been observed thus far.

\subsection{Ising-Coulomb Model}
The Ising-Coulomb (IC) model, in which the nearest neighbor ferromagnetic coupling spin is frustrated by long-range Coulomb interaction of relative strength $Q$, was first suggested as a model for the stripe phase behavior of high-temperature superconductors in two dimensions~\cite{emery:1993,low:1994}. It was also adopted as a generic coarse-grained description of microphase formation in systems with competing pair interactions in three dimensions~\cite{tarjus:1998,tarjus:2000,tarjus:2001}, and used to study the effect of dispersion forces on phase transitions in ionic systems~\cite{ciach:2001}. Although it is based on an Ising model, its Hamiltonian
\begin{equation}
H_{IC}=-J\sum_{\langle i,j\rangle}s_is_j+QJ\sum_{i>j}\frac{s_is_j}{r_{ij}},
\end{equation}
does not allow ferromagnetic ordering for any $Q>0$, i.e., an infinitesimally small Coulomb frustration is
sufficient to induce layering~\cite{tarjus:1998}. But by analogy with a Landau-Ginzburg model with frustration~\cite{emery:1993,low:1994,glotzer:1994,muratov:2002}, it is expected that any screening of the Coulomb interaction would move the onset of modulation to a finite $Q$~\cite{tarzia:2006,tarzia:2007}. Interestingly, the $Q\rightarrow\infty$ limit recovers the simple-cubic lattice restricted primitive model (LRPM) of Dickman and Stell at full occupancy~\cite{dickman:1999,stell:1999}.

The one-dimensional $T=0$ phase sequence is known to be made of equal length blocks of alternating
orientation~\cite{giuliani:2006}. In higher dimensions, though no rigorous demonstration exists,
layered phases of integer periodicity are also expected to be the ground state at low $Q$~\cite{tarjus:2000}. In that regime, an approximate mapping to a one-dimensional system seems reasonable. For sufficiently large $Q$,
two- and three-dimensional periodic structures, i.e., ``cylinders'' and ``clusters'', minimize the energy; and for
$Q>Q_N^0\approx 15.33$ 
antiferromagnetic N\'eel order is expected~\cite{tarjus:2000}.
Mean-field treatments~\cite{tarjus:2000,ciach:2001} and Monte Carlo simulations (for $Q<1$)~\cite{tarjus:2001}
describe the paramagnetic-modulated (PM) transition. 
Although the mean-field
results overestimate the transition temperature~\cite{tarjus:2000,ciach:2001}, the predictions are nonetheless quite similar
to the phase behavior obtained from simulations~\cite{tarjus:2001}. Because of the long-range isotropic Coulomb
interaction, the transition is ``fluctuation-induced'' first order for any $0<Q<Q_N$~\cite{brazovskii:1975,nussimov:1999}, and at low $Q$ the modulated phases melt at $T_c(Q)\sim T_c(0)-Q^{1/4}$, where $T_c(0)\approx 4.51$ is the 3D Ising simple cubic critical point~\cite{tarjus:1998}.
For $Q\geq Q_N$ the continuous paramagnetic-N\'eel (PN)
transition has Ising universality, and at high $Q$, the critical temperature $T_c(Q)\sim T_c(\infty) Q-T_c(0)$~\cite{tarjus:2000}, where the trivial linear dependence results from the choice of units and $T_c(\infty)\approx0.515$ is known from LRPM simulations~\cite{dickman:1999,almarza:2001}. A triple point connects the paramagnetic, modulated cluster, and antiferromagnetic N\'eel phases at $(Q_N,T_c({Q_N}))$, where
only the mean-field estimates $Q_N=36/\pi\approx 11.5$ and $T_c(Q_N)= 1.61$ are known~\cite{tarjus:2000}.
Within the modulated layered regime proper, phases spring out at the boundary between neighboring low-temperature ground states of integer periodicity~\cite{tarjus:2001}. The process is akin to the springing of phases between the antiphase and the ferromagnetic phase in the ANNNI model. The simulations in the layered regime capture the presence of these phases, but the use of a simulated-annealing approach in a strongly hysteretic regime is likely to bias the estimates for the transition temperatures~\cite{tarjus:2001}.



\section{Method}
\label{sect:method}
Monte Carlo simulations are used for determining the absolute free energy of the different modulated phases. The thermodynamic integration, the reference systems, and the Monte Carlo sampling details are presented in this section. 

\subsection{Thermodynamic Integration}
\begin{figure*}
\includegraphics[width=3.5in]{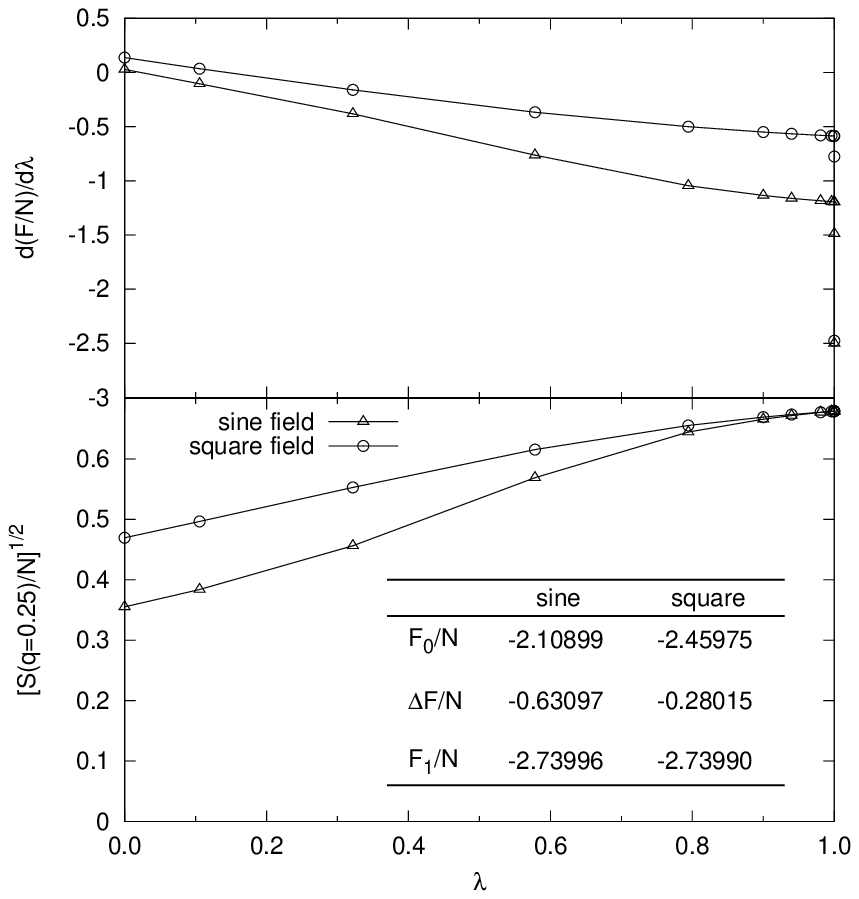}
\includegraphics[width=3.5in]{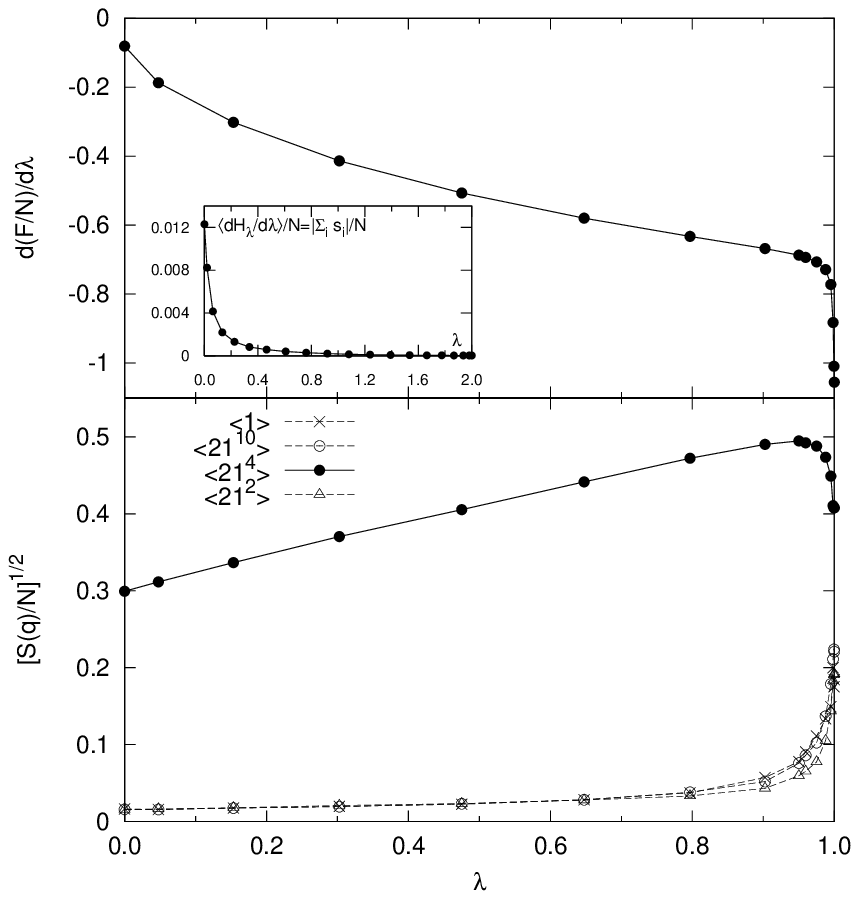}
\caption{(top left) Thermodynamic integration of the  ANNNI model
at $\kappa=0.7$ and $T=2.5$ for phase $\langle2\rangle$ using sinusoidal and square fields as reference. (bottom left) Change in the structure factor peak height
along the integration path and the free energy results for the two different references.
(top right) Thermodynamic integration of the IC model at
$Q=0.8$ and $T=1.06$ for phase $\langle21^4\rangle$. The integration curve from a fluctuating to a constant magnetization system is shown in the inset. (bottom right) The structure factor at different wavevectors demonstrates the preservation of the modulation along the integration path.} \label{fig:dfdlambda}
\end{figure*}

The free energy is obtained from Kirkwood thermodynamic integration~\cite{kirkwood:1935,frenkel:2002}, which
involves simulating a system with a Hamiltonian that couples a reference system Hamiltonian $H_0$ with that of the system of interest $H_1$
\begin{equation}
H_\lambda=(1-\lambda)H_0+\lambda H_1.
\end{equation}
For a given $\lambda$, the Helmholtz free energy $F_\lambda$ obeys
\begin{equation}
\frac{\partial F_\lambda}{\partial\lambda}=-\frac{T}{
Z_\lambda}\frac{\partial
Z_\lambda}{\partial\lambda}=\left\langle\frac{\partial
H_\lambda}{\partial\lambda}\right\rangle_\lambda,
\end{equation}
where $Z_\lambda$ is the canonical partition function and
$\left\langle\cdot\cdot\cdot\right\rangle_\lambda$ denotes a
canonical average under $H_\lambda$. The difference between the free energy
of system of interest $F_1$ and that of a known reference system $F_0$ at phase point $(T_0,\kappa_0)$ is thus
\begin{equation}
\begin{split}
 F_1(T_0,\kappa_0)-F_0(T_0,\kappa_0)&=\int_{0}^{1}\left\langle\frac{\partial
 H_\lambda}{\partial\lambda}\right\rangle_\lambda
 d\lambda\\
 &=\int_{0}^{1}\left\langle H_1-H_0\right\rangle_\lambda
 d\lambda.
\end{split}
\end{equation}
In order to obtain reliable numerical results, the integration
path from $\lambda=0$ to 1 must be reversible. No first order phase transition may take place along it. Our choice of reference system, which is key to the approach, is detailed
in the next subsection. 
The numerical integration is done by simulating the system at discrete $\lambda$ points chosen following
a Gauss-Lobatto scheme~\cite{abramowitz:1965}. Because of a rapid change in the integration curve as
$\lambda\rightarrow 1$, the latter part of the integral uses logarithmically spaced points that
are densely distributed near $\lambda=1$ (Fig.~\ref{fig:dfdlambda}). This adjustment is necessary for
accurately capturing the $z$-axis translational degree of freedom of the lattice, whose contribution is particularly
important in small systems.

In principle, one could investigate the $T$-frustration plane point by point,
but that would be computationally wasteful. The data collection is significantly accelerated by thermally integrating
to nearby temperatures $T_1$ or frustrations $\kappa_1$  (or, equivalently, $Q_1$) using a known state point $(T_0,\kappa_0)$ as reference by
\begin{equation}
\frac{F_1(T_1,\kappa_0)}{T_1}-\frac{F_1(T_0,\kappa_0)}{T_0}=\int_{T_0}^{T_1}\left(\frac{\partial F_1}{\partial
1/T}\right)_{\kappa_0} d(1/T)
\end{equation}
or
\begin{equation}
F_1(T_0,\kappa_1)-F_1(T_0,\kappa_0)=\int_{\kappa_0}^{\kappa_1}\left(\frac{\partial F_1}{\partial
\kappa}\right)_{T_0}d\kappa,
\end{equation}
where
\begin{equation}
\left(\frac{\partial F_1}{\partial
1/T}\right)_{\kappa_0}=\left\langle H_1\right\rangle_{\kappa_0}
\end{equation}
and
\begin{equation}
\left(\frac{\partial F_1}{\partial
\kappa}\right)_{T_0}=\left\langle\frac{\partial H_1}{\partial
\kappa}\right\rangle_{T_0}=\left\langle J\sum_{[i,j]_z}s_i s_j\right\rangle_{T_0}.
\end{equation}
In practice, the free energy results are
fitted with a polynomial of degree three or
four. The free energy at any point within a relatively short interval is then interpolated from the parameterized function.

\subsection{Reference System}
\label{sect:refsys}
In order to guarantee a reversible integration path, the reference system should reflect the symmetry
of the phase under investigation. A good reference system should also have a Hamiltonian $H_0$ whose
partition function $Z_0$ and free energy $F_0$ can be obtained analytically or at least with high numerical accuracy. For the lamellar phases observed on lattices with $N=L_xL_yL_z$ sites, we propose a reference that has decoupled spins under a $z$-axial periodically oscillating field $B(z)$ with amplitude $B_0$
\begin{equation}
H_0=-B_0\sum_{i=1}^Ns_iB(z_i),
\end{equation}
similarly to the periodic potential wells confining free particles used in Ref.~\onlinecite{mladek:2007}.
It trivially follows that in a system with fluctuating magnetization
\begin{equation}
\frac{F_0}{NT}=-\frac{1}{L_z}\sum_{z=1}^{L_z}\ln\left[2 \cosh\left( \frac{B_0B(z)}{T}\right)\right].
\label{eq:f0}
\end{equation}
The amplitude $B_0$ should be sufficiently strong to prevent layer melting and changes of layer periodicity
as the field is turned off, yet sufficiently weak to allow sampling of the integrand~\cite{footnote:1}. Fortunately, the relatively high free-energy barriers between neighboring modulated phases make phase transitions along
the integration path highly unlikely, even if sections of the path are formally metastable. Due to
the broken symmetry between the different coordinate axis, we can also, without loss of generality, similarly lock the lamellae in a specific orientation when initializing configurations for the IC model.

The applied field $B(z)$ needs not be the exact equilibrium profile of the modulated layers as long as the integration from $B(z)$ can be done reversibly. For instance, either square
or sinusoidal fields can be used as reference states for the study of modulated phases with integer
periodicity. The free energy results of both approaches agree with high accuracy (Fig.~\ref{fig:dfdlambda}). The equivalence also holds in the low-temperature regime,
where the ground state profile is more akin to a square well than to a pure sine function~\cite{selke:1979}.
Because sinusoidal fields are ``soft'' in the interlayer region, which helps averaging the layer fluctuations,
and because they provide a compact and efficient way to describe non-integer periodic lamellae, we use
\begin{equation}
B(z)=\sin(2\pi qz+ \phi_0),
\end{equation}
where a small phase angle $\phi_0$ is added to prevent the lattice sites from directly overlapping with the zeros of the field.

The IC model, which must remain charge neutral, requires that the reference partition function $Z_0$ be computed subject to a fixed magnetization constraint. In the infinite system limit this correction is negligible, but on a finite lattice it may affect transition temperatures.
For the paramagnetic phase, the reference system Hamiltonian $H_0$ with $B_0=0$ results in $F_0/N=-T\ln 2$ for an unconstrained system (Eq.~\ref{eq:f0}),
but the properly constrained reference system instead has
\begin{equation}
\frac{F_0'}{NT}=-\frac{1}{N}\ln{N \choose N/2},
\label{eq:f0zero}
\end{equation}
where ${N \choose N/2}$ is the binomial coefficient. For $N=12^2\times24=3456$ spins, the difference between the two results is $\sim0.001T$, which may be significant because of the small entropy differences between layered phases. Calculating $F_0'$
is not, however, as straightforward for $B_0\neq 0$. One has to define a thermodynamic integration path between fluctuating and constant magnetization systems
\begin{equation}
H_{\lambda}=H_0+\lambda\left|\sum_i s_i\right|,
\end{equation}
with $\lambda$ going from $0$ to $\infty$. In practice,
$\left\langle\frac{\partial H_{\lambda}}{\partial\lambda}\right\rangle=\left|\sum_i s_i\right|$ rapidly decays to zero
with growing $\lambda$, and therefore integrating to a finite $\lambda$ of order unity is sufficient.
The zero-magnetization free energy $F_0'$ is then obtained by adding the correction from thermodynamic
integration to $F_0$ from Eq.~\ref{eq:f0} (Fig.~\ref{fig:dfdlambda}). 
We note, however, that even in the small IC systems
studied here, the free energy corrections for different modulated phases are very similar for a given temperature. The
phase transitions are thus only imperceptibly affected by the shift.

\subsection{Monte Carlo Sampling}
We perform constant $T$ Monte Carlo (MC) simulations on a cubic lattice under periodic boundary conditions, using
$N=L_{x} L_{y} L_{z}=40^2\times240$ spins for the ANNNI model and $N=12^2\times24$ spins for the IC model,
unless otherwise noted. Ewald summation is used to compute the long-range Coulomb interactions in the
IC model~\cite{ewald:1921,frenkel:2002}. The phases studied have wave numbers $q=n/L_{z}$ with integer $n$'s,
which keeps modulations commensurate with the lattice. We initialize the modulated phases with a
sinusoidally varying spatial probability of the desired periodicity. The system relaxes to the equilibrium
spin profile for a given $xy$ plane
\begin{equation}
s_{xy}(z)\equiv\frac{1}{L_xL_y}\sum_{i\in xy}\langle s_{i}\rangle
\end{equation}
irrespectively of the initialization scheme as long as it has the correct periodicity. 

Basic MC sampling consists of single-spin flips for the ANNNI model. Spin exchanges, which enforce charge neutrality, are used for the IC model. Phase-space exploration gains in efficiency by complementing the basic sampling with iterations that take advantage of phase symmetry.
\begin{itemize}
\item For the modulated phases, layer swaps allow for the individual layer thickness to fluctuate while preserving the overall periodicity. 
    Multiple layer swaps are necessary to alter the periodicity and therefore even neighboring modulated phases are well separated in configuration space.
\item Near the PM transition, an anisotropic cluster algorithm for the ANNNI model~\cite{pleimling:2001} and a modified Wolff algorithm that considers the corrections from long-range interaction for the IC model~\cite{tarjus:2001} are used, in order to capture the strong fluctuations.
\item For systems with an applied external magnetic field, lattice drifts with respect to the field along the $z$ axis help sample the translational degrees of freedom.
\end{itemize}
For reference point integrations, up to $10^{5}$ MC moves ($N$ attempted spin flips or exchanges per move) are performed after $5\times10^{4}$ MC moves of preliminary equilibration. For the thermal and frustration integrations, only $10^{4}$ MC moves are necessary, because the free energy is not as sensitive to the accuracy of the integration slope as it is to its starting point, over the small $T$ intervals considered. In the vicinity of the critical transitions, we also use the multiple histogram algorithm, in order to obtain high precision results with a minimal amount of computations~\cite{ferrenberg:1991}. The method relies on reweighing the sampled configurations at a fixed temperature $T_0$, typically nearby $T_c$, by the Boltzmann factor difference
$e^{-(1/T-1/T_0)E}$ for results at neighboring temperatures $T$~\cite{newman:1999}. Our implementation uses a logarithmic summation scale, in order to avoid sum overflow in large systems~\cite{newman:1999}.


\section{Order and Critical Parameters}
\label{sect:order}
Structural order parameters help locate phase transitions, and are particularly important for the study of continuous and weakly first-order transitions in models studied here. The generalization and application of the study of critical and roughening transitions in modulated phases is presented in this section.

\subsection{Modulation Order Parameters}
\label{sect:orderparameter}
\begin{figure}
\includegraphics[angle=270,width=3.0in]{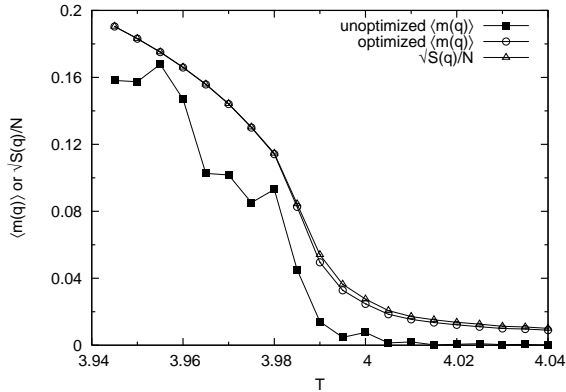}
\caption{Optimized and un-optimized $\langle m(q)\rangle$ and
$\sqrt{S(q)/N}$ for the ANNNI model at $\kappa=0.7$ and $q_c=0.1917$. The difference between $S(q)/N$
and $\langle m(q)\rangle^2$ gives $\chi(q)$ (Eq.~\ref{eq:susceptibility}).} \label{fig:mqsqannni}
\end{figure}
Functions of the Fourier spin density
\begin{equation}
\tilde{s}_q\equiv\sum_{i=1}^{N}s_{i}e^{i 2\pi qz_i}
\end{equation}
are natural choices for characterizing modulations in layered systems. The simplest of them, the generalized magnetization per spin, is defined analogously to the absolute magnetization in the Ising model~\cite{tarjus:2001}
\begin{equation}
\begin{split}
\langle m(q)\rangle &=\frac{1}{N}\sqrt{\langle \tilde{s}_{q}\rangle\langle
\tilde{s}_{-q}\rangle} \\
&=\frac{1}{N}\sqrt{\left\langle\sum_i s_i \cos(qz_i)\right\rangle^2 + \left\langle\sum_i s_i \sin(qz_i)\right\rangle^2 }.
\end{split}
\end{equation}
A direct use of $\langle m(q)\rangle$, however, causes problems in long simulations, because in principle it averages to zero as the
lattice drifts (Fig.~\ref{fig:mqsqannni}). Maximizing the real component of $\tilde{s}_{q}$ with respect to
a phase shift in the $z$ direction for each configuration before taking the thermal average resolves this issue.
In practice, we use a straightforward parabolic interpolation scheme~\cite{press:1992}.
Using the optimized version of $\langle m(q)\rangle$, even in simulations that
are too short for the system's periodicity to completely diffuse, significantly improve the data quality
(Fig.~\ref{fig:mqsqannni}). Only quantities based on optimized $\tilde{s}_{q}$ are therefore used in the rest
of this work. The generalized magnetization decays $\langle m(q)\rangle\sim(T_c-T)^{\beta}$ with critical exponent $\beta$, but the decay properties are not ideal for numerically detecting critical temperatures in finite systems.

\begin{figure}
\includegraphics[width=4.5in]{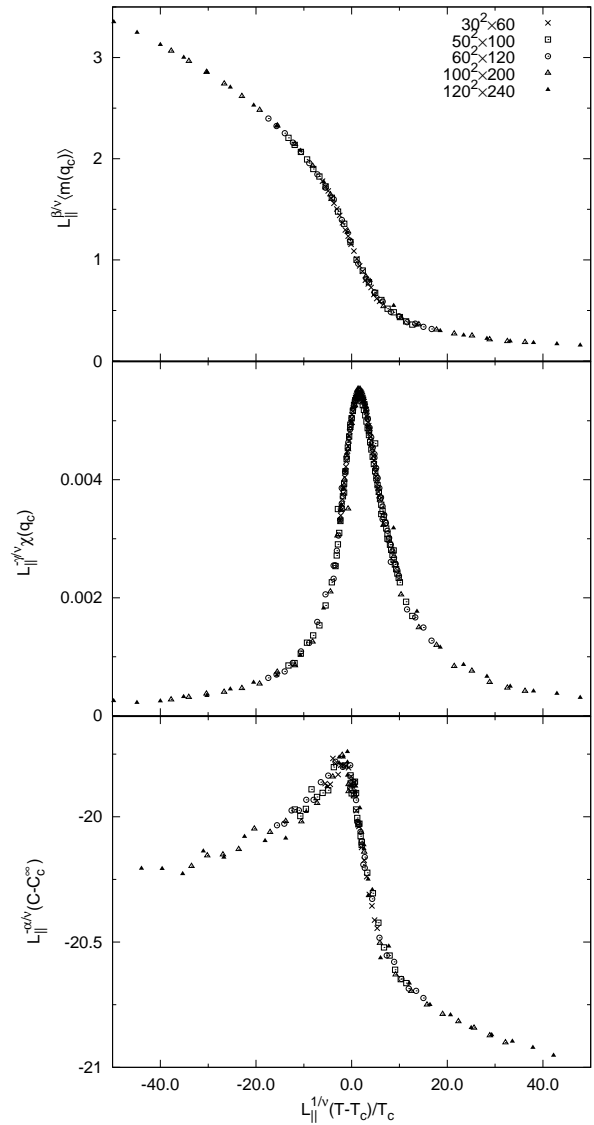}
\caption{Finite-size scaling of the magnetization (top), susceptibility (middle) and heat capacity (bottom)
of the ANNNI model at $\kappa=0.8$ and $q_c=0.2$, 
using $\nu=0.66$, $\gamma=1.32$, $\beta=0.34$, $\alpha=-0.01$, and
$C_c^\infty=20$. 
}
\label{fig:finitesizeK0.8}
\end{figure}

The next higher magnetization moment, the $z$-axial static structure factor, is similar to the equivalent liquid-state quantity
\begin{equation}
\begin{split}
N S(q)& \equiv\langle\tilde{s}_q\tilde{s}_{-q}\rangle=N^2\langle m^2(q)\rangle\\
&=\left\langle \left|\sum_i s_i \cos(qz_i)\right|^2+ \left|\sum_i s_i \sin(qz_i)\right|^2\right\rangle,
\end{split}\label{eq:structurefactor}
\end{equation}
where $\langle m^2(q)\rangle$ is the second moment of the magnetization~\cite{stanley:1977a,selke:1979}. Both $S(q)$ and its normalized version $\sqrt{S(q)/N}=\sqrt{\langle m^2(q)\rangle}$~\cite{tarjus:2001} grow upon cooling and are maximal at the wave number $q_c$ of the first modulated phase below $T_c$. The monotonically increasing $S(q)$ is, however, ill-suited for detecting the PM transition in simulations, because, like $\langle m(q)\rangle$, it does not give a clear visual signature of $T_c$. The generalized susceptibility
\begin{align}
T \chi(q)&\equiv\frac{1}{N}(\langle \tilde{s}_q\tilde{s}_{-q}\rangle-
\langle\tilde{s}_{q}\rangle\langle \tilde{s}_{-q}\rangle)\nonumber\\
&=N\langle
m^2(q)\rangle-N\langle m(q)\rangle^2,
\label{eq:susceptibility}
\end{align}
i.e., the second cumulant of the magnetization, does not suffer from this caveat. It
indeed diverges on both sides of the transition $\chi(q_c)\sim|T-T_c|^{-\gamma}$ with critical exponent $\gamma$, as would $\chi(0)$ in the Ising model, and was used in our previous study~\cite{zhang:2010} (Fig.~\ref{fig:finitesizeK0.8}). Directly correcting for finite-size effects, however, results in a high-sensitivity of the transition location to simulation noise. The Binder cumulant route is more convenient for detecting $T_c$, because its value at the critical point $U_4^*$ is straightforwardly insensitive to scaling the system size~\cite{binder:1981,landau:2000}. For layered phases, a generalization of the expression
\begin{equation}
U_4(q)=1-\frac{\langle m^4(q)\rangle}{3\langle m^2(q)\rangle^2}
\end{equation}
in terms of the second and the fourth $\langle m^4(q)\rangle=\langle
\tilde{s}_q^2\tilde{s}_{-q}^2\rangle/N^4$ $q$-modulated magnetization moments provides the necessary information.

\begin{figure}
\includegraphics[width=3.4in]{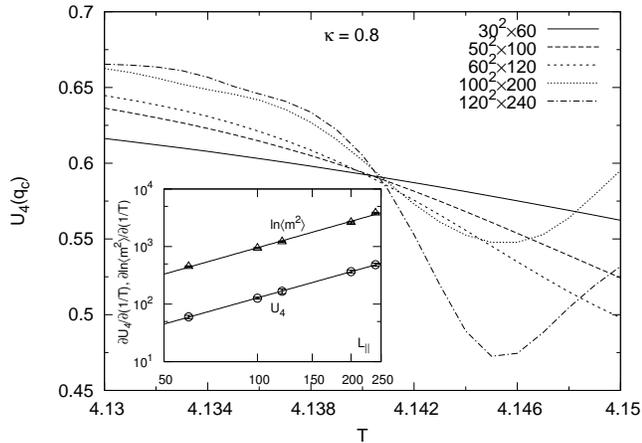}
\caption{Binder cumulant of the ANNNI model at $\kappa=0.8$ and $q_c=0.2$.
The curves, which intersect at the critical temperature $T_c=4.141$, monotonically decrease with $T$. The limited validity regime of histogram reweighing results is here responsible for the non-monotonicity.
(Inset) Finite-size scaling analysis of the peak of the derivative of $U_4(q_c)$ and $\ln \langle m(q_c)^2\rangle$.
The logarithm scales as $1/\nu$, which here gives $\nu=0.66(2)$.}
\label{fig:U4K0.8}
\end{figure}

Because of the anisotropy of the modulated phases, it is useful to review how breaking isotropy may affect critical properties. In a system of dimensions parallel $L_{\parallel}\equiv L_z$ and perpendicular $L_{\perp}\equiv L_x=L_y$ to the modulation propagation, the correlation length $\xi$ may diverge with different critical exponents
\begin{equation}
\xi_{\parallel}\sim |T-T_c|^{-\nu_{\parallel}},\\~~~~
\xi_{\perp}\sim |T-T_c|^{-\nu_{\perp}}.
\end{equation}
The critical Binder cumulant $U_{4}^*= U_4(q_c,T_c)$ is then invariant for a fixed ratio $L_{\parallel}/L_{\perp}^{\nu_{\parallel}/ \nu_{\perp}}$ (Fig.~\ref{fig:U4K0.8})~\cite{binder:1989,footnote:2}. At a uniaxial Lifshitz point, such as in the ANNNI model~\cite{pleimling:2001}, $\nu_{\parallel}\simeq \frac{1}{2}\nu_{\perp}$~\cite{hornreich:1975,diehl:2000}. For the PM transition, at $\kappa>\kappa_L$, we also consider the possibility of anisotropic critical behavior. Although a direct determination of $\nu_{\parallel}/ \nu_{\perp}$ is numerically difficult, our indirect finite-size study of systems with a fixed ratio $L_{\parallel}/L_{\perp}=2$ shows that $U_4^*$ does not vary at the PM transition (Fig.~\ref{fig:U4K0.8}). This observation suggests that $\nu_{\parallel}/ \nu_{\perp}\approx 1$, i.e., $\xi_{\parallel}$ and $\xi_{\perp}$ diverge with the same critical exponent $\nu_\parallel=\nu_\perp\equiv\nu$ at the PM transition. Critical anisotropy is thus neglected in the rest of this study.

Binder cumulants also allow to independently determine the critical exponent $\nu$ using the peak value of the derivative of $U_4$
\begin{equation}
\ln \left(\frac{\partial U_4}{\partial1/T}\right)_{\mathrm{max}}=\frac{1}{\nu}\ln L+\mathrm{constant}.
\end{equation}
A similar relation for the structure factor gives~\cite{ferrenberg:1991,landau:2000}
\begin{equation}
\ln \left(\frac{\partial\ln\langle m^2(q)\rangle}{\partial1/T}\right)_{\mathrm{max}}=\frac{1}{\nu}\ln L+\mathrm{constant}.
\end{equation}
The system size $L$ in the scaling relation can be either $L_{\parallel}$ or $L_{\perp}$ as long as the ratio $L_{\parallel}/L_{\perp}$ is fixed. For an isotropic critical point, as long as the dimensions
are rescaled by the same factor, the form of the collapse and the critical exponents remain unchanged (Fig.~\ref{fig:U4K0.8}).

Once $\nu$ and $T_c$ are obtained, the critical exponents $\beta$ and $\gamma$ can more easily be determined through
finite-size scaling~\cite{landau:2000}. The quantities $L^{\beta/\nu}\langle m(q_c)\rangle$ and $L^{-\gamma/\nu}\chi(q_c)$
overlap for different system sizes, when drawn as a function of the scaled temperature $L^{1/\nu}(T-T_c)/T_c$.
The heat capacity can also be similarly rescaled, but only if $C$ diverges at $T_c$, as at transitions with Ising universality.
For a transition with XY-universality, for which $\alpha=-0.01$, $C$ peaks at a finite value $C^{\infty}_c$ in
the infinite system size limit. The proper scaling relation is then $L^{-\alpha/\nu}(C-C^{\infty}_c)$~\cite{gottlob:1993,schultka:1995}. Rescaling the heat capacity curves for such a small $\alpha$ is, however,
subject to sizable numerical errors~\cite{gottlob:1993}. The hyperscaling relation $2-\alpha=3\nu$ is used instead to determine $\alpha$~\cite{campostrini:2001}.

\subsection{Interfacial Roughening}
\begin{figure}
\includegraphics[width=\columnwidth]{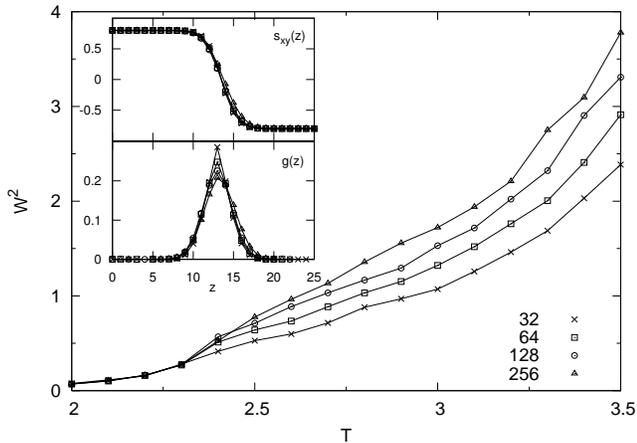}
\caption{The roughening transition in the ferromagnetic regime at $\kappa=0.2$ using systems of size $L^2\times26$ is found at $T_R=2.35(5)$. (insets) Average magnetization profile and gradient at $T=3.5$.}
\label{fig:rough_sample}
\end{figure}

In the Ising ferromagnetic regime, even though the correlation length isotropically diverges at the critical point,
the interface between two regions of opposite magnetization presents a roughening transition $T_R$ at roughly half the critical temperature~\cite{weeks:1973,swendsen:1977,mon:1990,berera:1993}. Below $T_R$ the interface is localized, while above $T_R$ its width diverges logarithmically with surface area. In simulation, an interface is created within the bulk by using an antiperiodic boundary condition along the $z$ direction~\cite{weeks:1973}, and the transition can be localized by finite-size analysis (Fig.~\ref{fig:rough_sample}).
Because modulated phases intrinsically present a series of interfaces between regions of opposite magnetization, it is also interesting to consider whether these interfaces roughen or not with temperature. It has been suggested that they too should logarithmically diverge~\cite{levin:2007,brazovskii:1975}. If that were the case, it is possible that interlayer fluctuations at temperatures
between $T_R$ and $T_c$ could participate in phase branching and the formation of equilibrium incommensurate structures in the large system limit~\cite{bak:1982}. A generalization of the simulation approach is here necessary.

The variance of the interface position $z$
\begin{equation}
W^2\equiv\langle(z-\langle z\rangle)^2\rangle
\label{eq:w2}
\end{equation}
measures the fluctuations of the interface location, and is expected to diverge logarithmically with system size
for fixed $T>T_R$
\begin{equation}
W^2\sim \ln L_{\perp}.
\end{equation}
For $T< T_R$, $W^2$ should have an even weaker system size dependence.
In practice, the average in Eq.~\ref{eq:w2} is taken over the normalized magnetization gradient
\begin{align}
g(z)&=\frac{(ds_{xy}(z)/dz)}{\int(ds_{xy}(z)/dz) dz}\\
&=\frac{s_{xy}(z+1)-s_{xy}(z)}{\sum_z[s_{xy}(z+1)-s_{xy}(z)]},
\end{align}
which serves as weight function~\cite{burkner:1983}. The equilibrium profile $s_{xy}(z)$ is obtained by aligning instantaneous profiles to correct for lattice drift before averaging (Fig.~\ref{fig:rough_sample}).
For the modulated regime, where multiple interfaces are present, layers within half a period of the interface $i$, i.e., layers whose $z$ coordinates belong to a set $I$, are grouped together in the variance calculation
\begin{equation}
W_i^2=\sum_{z\in I} z^2g(z)-\left(\sum_{z\in I} zg(z)\right)^2,
\end{equation}
and the results for the various interfaces are averaged at the end.

\section{Results and Discussion}
\label{sect:results}
Assembling the results from the various observables obtained from the computational techniques
provides a clearer understanding of the equilibrium phase behavior of the ANNNI and IC models.
In this section, we concentrate on the properties of the modulated regime. 

\subsection{Phase Transitions between Layered Phases}
\begin{figure*}
\includegraphics[width=\columnwidth]{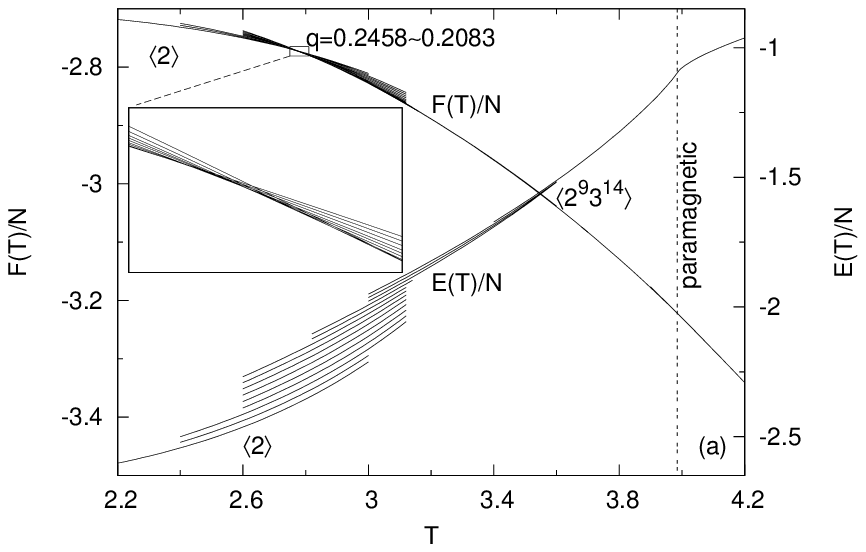}
\includegraphics[width=\columnwidth]{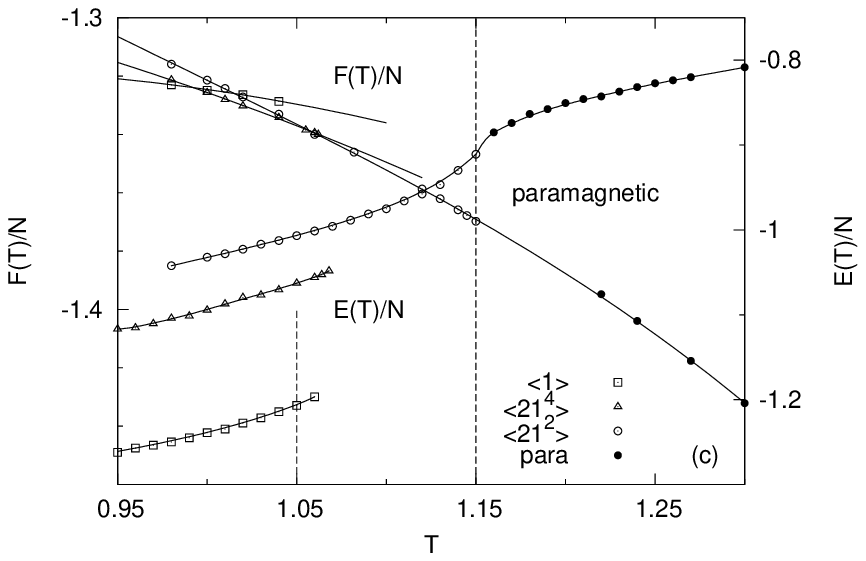}
\includegraphics[width=\columnwidth]{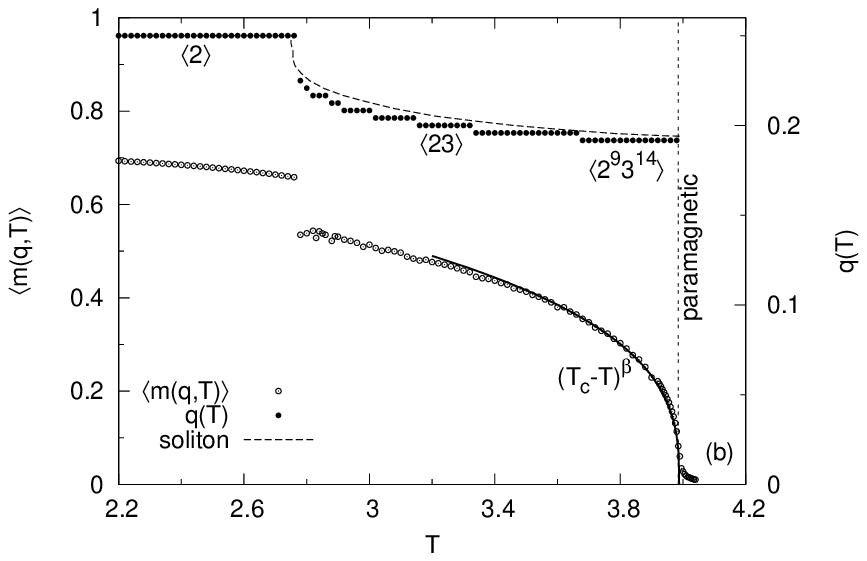}
\includegraphics[width=\columnwidth]{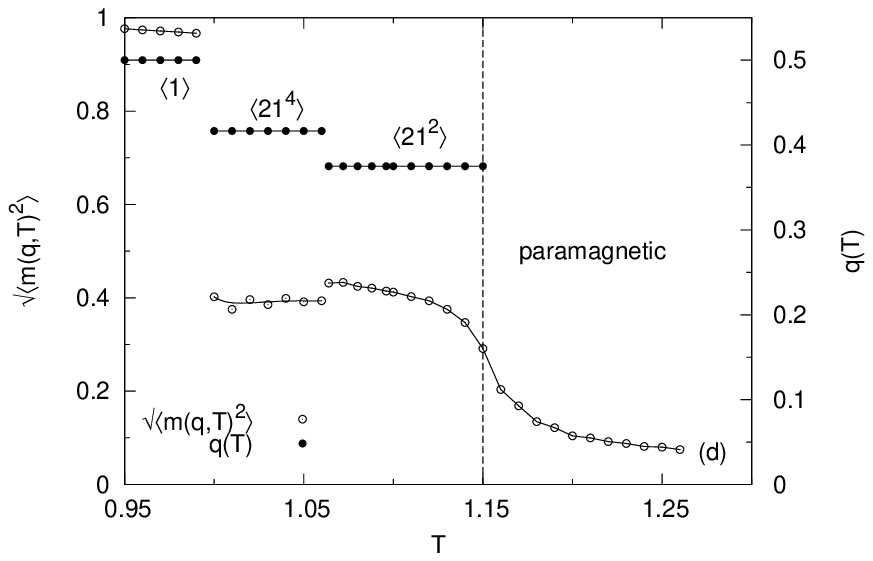}
\caption{(a) Energy and free energy results for the ANNNI model at $\kappa=0.7$ for modulations ranging from phase $\langle 2\rangle$ to phase $\langle 2^93^{14}\rangle$ at melting. The PM transition $T_c=3.988$ (vertical dashed line) is obtained from $U_4^*$.
(b) Under the same conditions as (a), equilibrium devil's staircase compared with the rescaled soliton result (dashed line), and $\langle m(q)\rangle$ compared with the power-law decay form with $\beta=0.34$ obtained from finite-size scaling (solid line). Note that in the low-temperature limit, the square profile of phase $\langle 2\rangle$ gives $\langle m(q)\rangle\rightarrow 2^{-1/2}$.
(c) Energy and free energy results for the IC model at $Q=0.8$  for phases $\langle1\rangle$, $\langle21^4\rangle$ and $\langle21^2\rangle$, and the paramagnetic phase.
The PM transition $T_c=1.15(1)$ is obtained from  $C$. The short vertical dashed line indicates the transition temperature between phases $\langle 1\rangle$ and $\langle 21^2\rangle$, obtained from simple annealing (see text)~\cite{tarjus:2001}.
(d) Under the same conditions as (c), devil's staircase and normalized structure factor. Note that in the low-temperature limit, the profile of phase $\langle 1\rangle$ gives $\sqrt{\langle m(q)^2\rangle}\rightarrow 1$.}
\label{fig:ft}
\end{figure*}

\begin{figure}
\includegraphics[width=\columnwidth]{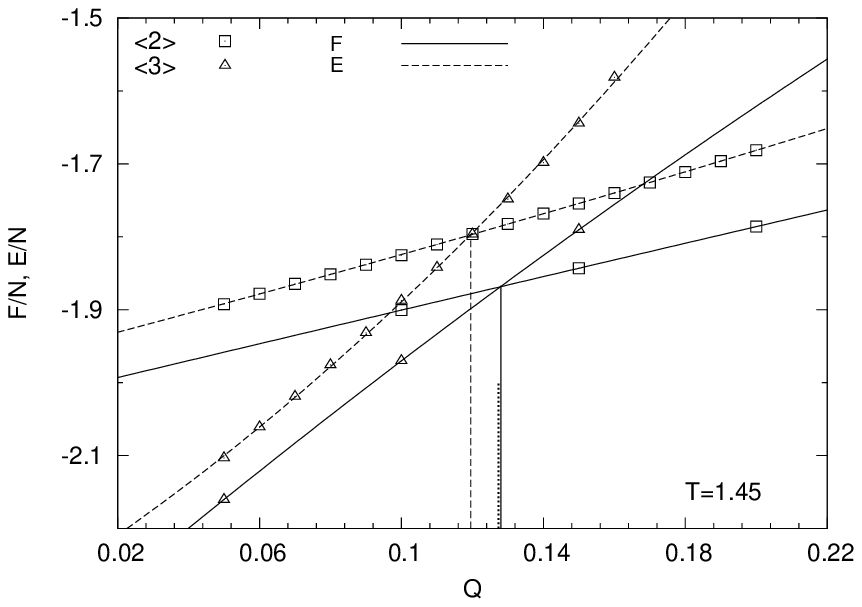}
\caption{Energy and free energy curves of the IC model for  phases $\langle3\rangle$ and $\langle2\rangle$ at $T=1.45$.
The phase boundary from free energy calculation (solid lines) agrees
with the $T=0$ mean field prediction (dotted line) and is different from the energy inversion (dashed lines)~\cite{tarjus:2000}.}
\label{fig:lowT}
\end{figure}

The size of the energy gap between neighboring phases with $q$'s commensurate with
the simulation box reflects the limited and constrained choice of modulations realizable on a finite periodic
lattice (Fig.~\ref{fig:ft}). In an infinite periodic system, where all rational modulations are valid but irrational
$q$'s are excluded, this gap would be infinitely small because rational numbers are dense on the real
axis~\cite{bak:1981,rasmussen:1981,selke:1988}. The smooth and extended energy curves for the different modulations
are also characteristic of strongly metastable phases (Figs.~\ref{fig:ft} and \ref{fig:lowT}). The high free-energy barriers between layers of differing
periodicity result in phases that are sufficiently long-lived to persist throughout the entire simulation,
if the $L_x L_y$ cross-section is large enough. Different phases can be observed at a given temperature
and frustration, depending on how the system is initialized.  For smaller cross sections, however, the reduced
number of spins involved in changing the periodicity lowers these transition barriers.
For a fixed system size, although a longer $L_z$ allows the study of more modulated phases, the need to
keep these phases stable limits the maximal aspect ratio $L_\parallel:L_\perp$ of the simulation lattice. In practice, the selected ratio must
balance these competing demands. A microscopic understanding of the transition
mechanism between layered phases of different periodicity is still incomplete~\cite{selke:1979}, but we empirically find that a size ratio of 2:1 is generally sufficient.

The crossing of free-energy curves of neighboring modulated phases identifies the transition temperature. Using this approach side steps the hysteresis that otherwise afflicts annealing approaches, and results in a more accurate depiction of the modulated regime than had previously been
obtained~\cite{selke:1980,rasmussen:1981,tarjus:2001,murtazaev:2009}. For the IC model, for instance, the free energy calculations
 locate the phase transitions at temperatures at least $10\%$ lower than reported in
Ref.~\onlinecite{tarjus:2001}, where the the system was prepared in the $T=0$ ground state and studied by simulated annealing. At $Q=0.8$,
we can even identify a commensurate modulated phase $\langle21^4\rangle$ that was entirely missed by the annealing study.
The other possibly missed commensurate phase $\langle21^{10}\rangle$ is, however, unstable here as well, presumably because of finite size effects
(Fig.~\ref{fig:ft}). Qualitatively similar results are obtained for $Q=0.144$ and $Q=0.17$  (not shown). 

By integrating over frustration at
low temperature we can also identify the boundary between phases of integer periodicity in the IC model.
The results at finite temperatures agree very closely with the $T=0$ energy derived transitions. The location of the free energy cross over between phases $\langle 1\rangle$ and $\langle 2\rangle$ (not shown) as well as between phases $\langle 2\rangle$ and $\langle 3\rangle$ is only mildly affected by temperature (Fig.~\ref{fig:lowT}). The thermal fluctuations produce a similar free energy shift of both phases, which leaves the $Q$ location of the transition unchanged. We thus expect similar results at other phase $\langle n\rangle$-$\langle n+1\rangle$ transitions.

The PM transition is not accurately obtained by direct free energy comparisons for either systems. For the ANNNI model, the continuous transition is best studied through the specialized tools of critical phenomena (see below). But even for the IC model, the fluctuation-induced first-order transition does not lead to sufficiently high free energy barriers
for noticeably supercooling the paramagnetic phase in such a small system. The system instead rapidly freezes into a modulated phase below the transition and shows only a minimum of hysteresis. As a result, the transition identified from the heat capacity peak by annealing in Ref.~\onlinecite{tarjus:2001} is equivalent to what is obtained here. A more careful system size dependence study would be necessary to refine the transition estimate. 

\subsection{Devil's Staircase and Order Parameter}
The equilibrium wave number obtained from the free energy results displays the characteristic devil's staircase~\cite{bak:1980}.
The stability regime of a given modulated phase stretches over an ever smaller $T$ range upon cooling.
For the ANNNI model, the predicted truncation of the sequence before reaching the antiphase makes the
staircase ``harmless''~\cite{fisher:1987}, but the simulated system size is here insufficient to distinguish this
scenario from the infinite ``devil's last step'' sequence in which no commensurate phase is missed~\cite{selke:1988,fisher:1987}. The overall shape of the
decay can, however, be compared with the soliton theory prediction~\cite{bak:1980}. Though the soliton does not
correctly capture the PM transition temperature, once $T$ is linearly rescaled to make $T_c$ coincide,
the agreement is fairly good (Fig.~\ref{fig:ft}).

The equilibrium generalized magnetization behaves similarly to its $q=0$ version in the Ising model. For the ANNNI model around $T_c$, the quantity grows monotonically upon cooling. It continuously increases at first, but upon reaching the antiphase it jumps discontinuously. In the antiphase region the magnetization profile tends toward a periodic square, whose profile structure is only partially captured by a simple sinusoidal function.
For the IC model, the renormalized structure factor, which is indistinguishable from $\langle m(q)\rangle$ at low temperatures, is also not an ideal order parameter. When the system changes from
phase $\langle21^{2}\rangle$  to $\langle21^{4}\rangle$, for instance,
the peak height actually goes down. Here again, the modulation profile is not well captured by a
simple sinusoidal function. The inclusion of higher order harmonics might better detect growing order upon cooling.


\subsection{Modulation at Melting}
\begin{table*}
\caption{Critical parameters of the ANNNI model for $\kappa\leq\kappa_L=0.270(4)$ obtained by finite-size scaling of cubic systems with $L=16,~32,~40,~64,~80$ for $\kappa=0.1$ and $0.2$ and from previous simulations~\cite{kaski:1985,pleimling:2001}. Ising values are given for reference. The uncertainty on $T_c$ and $q_c$ from the HT series expansion results from the Pad\'e approximant method~\cite{oitmaa:1985}. At $\kappa_L$, $\nu_\parallel$ is reported. The starred $^*$ $\alpha$ results are obtained from the hyperscaling relation $3\nu=2-\alpha$ (or $\alpha+2\beta+\gamma=2$ for $\kappa=0.24$).}
\begin{tabular}{c c c c c c c c}
\hline
$\kappa$ &Ising &0.1 &0.15~\cite{kaski:1985} &0.2 &0.24~\cite{kaski:1985} &0.265~\cite{kaski:1985} &0.270~\cite{pleimling:2001}\\
\hline
$T_c^{\mathrm{MC}}$ & 4.512 & 4.265(1) &4.15(2) & 3.987(1)&3.86(2) & 3.77(2) & 3.7475(5)\\
$T_c^{\mathrm{HT}}$ & 4.51(2) & 4.26(2)&4.13(2) & 3.98(2) &3.85(2) & 3.76(2)& 3.75(2)\\
$\nu$  &0.63  &0.62(1) & 0.61(3) & 0.62(2) & --&0.51(4) &0.33(3)~\cite{kaski:1985}\\
$\alpha$ &0.11 &$0.14(3)^*$ & $0.17(9)^*$ &$0.14(6)^*$ & $0.28(12)^*$ & $0.47(12)^*$ & 0.18(2)\\
$\beta$ &0.34 & 0.31(2)& 0.30(3)& 0.31(3)&0.23(3)& 0.19(2)& 0.238(5)\\
$\gamma$&1.24 &1.25(2) & 1.20(6)& 1.23(3)& 1.26(6)& 1.40(6) & 1.36 (3)\\
\hline
\end{tabular}
\label{table:criticalferro}
\end{table*}

\begin{table*}
\caption{See Table~\ref{table:criticalferro} for details. Critical parameters of the ANNNI model for $\kappa>\kappa_L$ obtained by finite-size scaling of systems with $L_z=60,~120,~150,~180,$ and $240$ at $\kappa=0.522$, $L_z=120,~240,$ and $360$ at $\kappa=0.7$, $L_z=60,~100,~120,~200,$ and $240$ at $\kappa=0.8$, $L_z=60,~120,~180,$ and $240$ at $\kappa=2.0$, and from previous simulations~\cite{rotthaus:1993,selke:1979}. XY values are given for reference.}
\begin{tabular}{c c c c c c c c}
\hline
$\kappa$ & XY~\cite{campostrini:2001}& 0.5~\cite{rotthaus:1993} &0.522 &0.6~\cite{selke:1979} & 0.7 &0.8  &2.0 \\
\hline
phase & --& $\langle 3\rangle$ & $\langle 3\rangle$ & $\langle2^93^{34}\rangle$ & $\langle2^93^{14}\rangle$ & $\langle 23\rangle$ & $\langle2^63 \rangle$ \\
$q_c^{\mathrm{MC}}$ &-- & 0.17(3) & 0.167(4) & 0.18(3) & 0.192(4) &0.200(4) &0.233(4)  \\
$q_c^{\mathrm{HT}}$ &-- & 0.162(1)& 0.167(1)&0.180(1)  & 0.192(1) & 0.200(1) &0.232(1) \\
$q_c^{\mathrm{MF}}$ & -- & 0.1667&0.1705 &0.1816 & 0.1919& 0.1994&0.2301 \\
$T_c^{\mathrm{MC}}$ & 2.202 & 3.6(1) &3.723(1) &3.82(3) & 3.988(1) &4.141(1) & 5.796(1) \\
$T_c^{\mathrm{HT}}$ &  2.202~\cite{adler:1993}   & 3.67(2)&3.72(2)  & 3.81(2)& 3.99(3) &4.14(1) &5.79(1)\\
$\nu$ & 0.67 & -- & 0.66(2)& -- & 0.67(4) &0.66(2) & 0.67(3)\\
$\alpha$ & -0.01 & -- & $0.02(6)^*$ & -- & $-0.01(9)^*$ & $0.02(6)^*$ &$-0.01(9)^*$ \\
$\beta$ &0.35 & --& 0.35(2) & -- & 0.35(3)& 0.34(3) &0.36(2) \\
$\gamma$ &1.32 & --& 1.30(4) & -- &1.32(8)   &1.33(4) & 1.32(6) \\
\hline
\end{tabular}
\label{table:criticalmodulated}
\end{table*}

\begin{figure}
\includegraphics[width=3in]{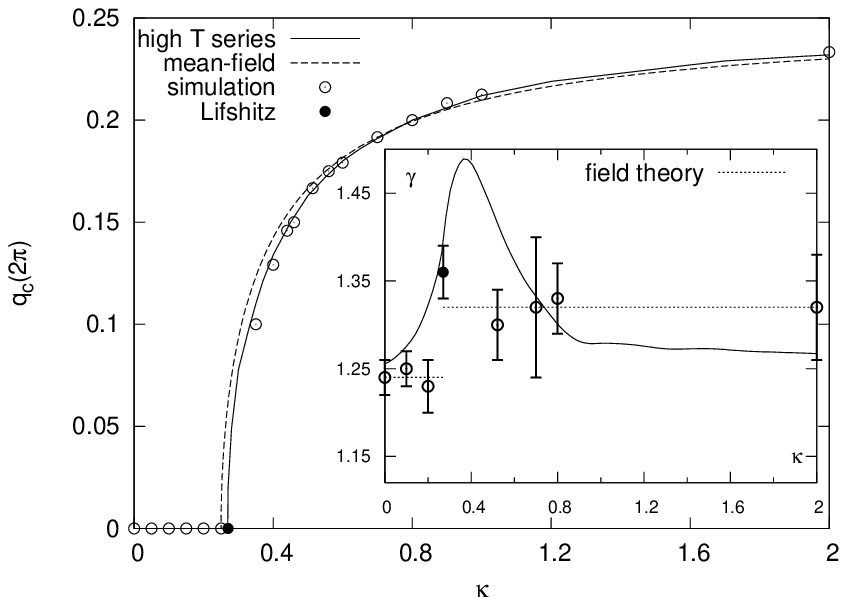}
\includegraphics[width=3in]{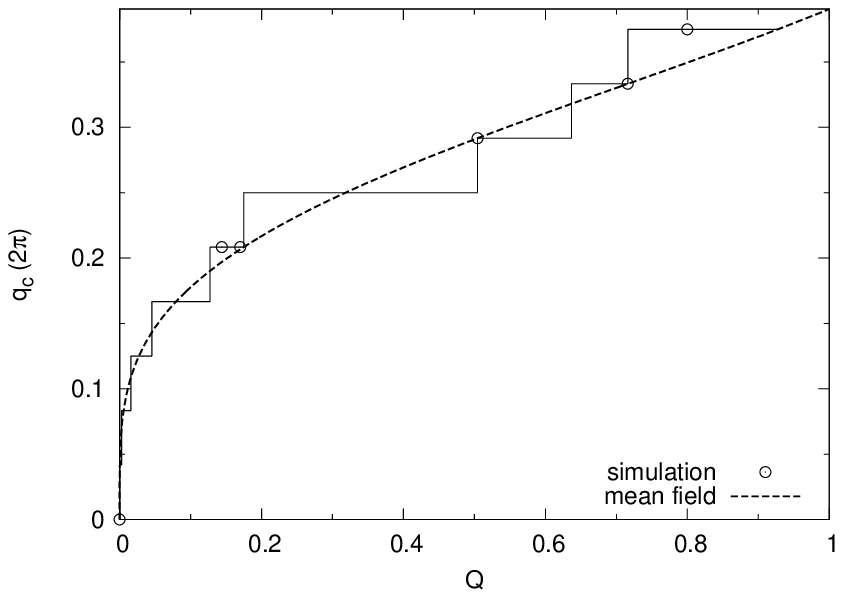}
\caption{Simulation wave number periodicity at melting $q_c$ for different frustration strengths of the ANNNI (top) and the IC (bottom) models superimposed with theoretical predictions. The solid line for the IC model captures the accessible wave number for $L_z = 24$. (top inset) The critical exponent $\gamma$ obtained by finite-size scaling is compared with the high-temperature
series expansion~\cite{oitmaa:1985}, and the field-theory predictions for the exponent~\cite{garel:1976,droz:1976}. The Lifshitz information is taken from Ref.~\onlinecite{pleimling:2001}.}
\label{fig:qc}
\end{figure}

The periodicity of the modulated phase at the PM transition $q_{c}(\kappa)$ (or $q_{c}(Q)$) is remarkably insensitive
to the theoretical approach used for capturing its behavior. In the ANNNI model, the agreement between
simulation results, mean-field theory~\cite{selke:1984}, HT series expansion~\cite{stanley:1977a,oitmaa:1985},
and the critical scaling near the Lifshitz point
\begin{equation}
q_c\sim|\kappa-\kappa_L|^{\beta_{l}},
\end{equation}
using either the critical exponent from series expansion $\beta_l=0.5\pm0.05$~\cite{stanley:1977a,oitmaa:1985} or
from renormalization group $\beta_l=0.514$~\cite{shpot:2001},
is very good (Fig.~\ref{fig:qc}). 
The similarity of the
RG critical exponent with the mean-field value further suggests that the dependence of microphase periodicity on
frustration is much easier to capture than the transition temperature. The free energy correction due to fluctuations is likely similar for neighboring layered phases.

For 
the IC model, the mean-field prediction for the continuously changing $q_c$ is also within the simulation accuracy
in the layered regime (Fig.~\ref{fig:qc}), but the relatively small lattice size limits quantitatively assessing the
theoretical predictions. In the high $Q$ regime, where
a N\'eel-paramagnetic-modulated phase triple point is expected, our coarse simulation estimate $Q_N\approx15.8$ clearly differs from the mean-field prediction $11.5$~\cite{tarjus:2000}. A similarly large deviation between the theoretical prediction and the direct calculation is also observed at $T=0$, where $Q_N^0=9.549$ and 15.33, respectively~\cite{tarjus:2000}. Both those differences can mostly, and possibly completely, be explained by the low accuracy of the lattice Fourier transform in the large $Q$ limit, where modulated phases of small domains form~\cite{tarjus:2000}.
Note, however, that the critical nature of the $Q_N$ point, which depends on the properties of the modulated-N\'eel transition, could also impact its location. If it is a bicritical point, fluctuations could result in larger deviations from the mean-field predictions. A generalization of the free energy simulation approach to other modulated geometries should be able to resolve this question, but is beyond the scope of this work.


\subsection{ANNNI Critical Behavior}
The critical properties of the ANNNI model have been extensively studied using high-temperature (HT) series expansion~\cite{stanley:1977a,oitmaa:1985,mo:1991}. For instance, critical temperatures can be estimated by resumming truncated series with Pad\'e approximants~\cite{oitmaa:1985,stanley:1977b}. The arbitrariness of selecting the Pad\'e order results in a range of estimates (Tables~\ref{table:criticalferro} and~\ref{table:criticalmodulated}). The values of $T_c$ obtained from finite-size scaling quantitatively agree with these estimates, and are an order of magnitude more precise than both the HT series results and previous simulation estimates~\cite{kaski:1985,rotthaus:1993}.

The critical exponents from the HT series expansion, however, only qualitatively agree with the simulation results. Finite HT series can only smoothly approximate changes in critical behavior, but the critical exponents change discontinuously on both side of the Lifshitz point. The HT series results are a continuous approximation of that singularity. Field theory arguments suggest, however, that the critical exponents should have  Ising universality below the Lifshitz point, XY above the Lifshitz point~\cite{garel:1976}, and uniaxial Lifshitz universality at the Lifshitz point~\cite{diehl:2000}. The Ising~\cite{kaski:1985}  and Lifshitz point~\cite{pleimling:2001} predictions have been previously confirmed by Monte Carlo simulations, but above the Lifshitz point the model's behavior is not so clear.  In particular, the HT series results for $\gamma$ at large $\kappa$ undershoot the XY exponent value. In the words of Ref.~\onlinecite{oitmaa:1985}, at high $\kappa$ ``a puzzling and unexplained feature [of the HT series expansion results] is the apparent decrease of $\gamma$ to something like the Ising value.'' Later similar studies did not quite resolve this question, and even suggested that a different type of universality might be observed beyond $\kappa\approx 2$~\cite{mo:1991}. Our earlier simulation results did not provide a clear resolution of this issue either, because of limited system sizes and insufficient averaging in the critical region~\cite{zhang:2010}. Simulation of larger systems using the multiple histogram method, however, lifts any remaining ambiguity. The critical exponent results support a XY universality of the transition for all $\kappa>\kappa_L$ studied. The values of $\nu$ and $\gamma$ agree with each other and with the XY values, and are often significantly different from the Ising exponents, despite the relatively large error bars (Table~\ref{table:criticalmodulated} and Fig.~\ref{fig:qc}). The finite-size scaling of $C$ using $\alpha$ derived from hyperscaling relations further supports the agreement (Fig.~\ref{fig:finitesizeK0.8}). These observations, however, shed some doubt on the validity of the predicted transition at $\kappa\approx2$. 

The XY universality of the PM transition can be understood from the similarity between its two-component order
parameter and that of the XY model~\cite{hornreich:1980,li:1989}. 
A mean-field picture for the order parameter of the ANNNI model (Eq.~\ref{eq:structurefactor}) suggests that a spin $i$ in the modulated phase can be thought of evolving within a magnetization profile of periodicity $q$
formed by all the other spins. A change of the average local magnetization at position $i$ is equivalent to shifting the phase angle $qz_i$ with respect
to that profile. Note that the isotropic nature of the critical PM transition suggests that pairs of spins parallel and perpendicular to the $z$ axis are equivalently correlated, i.e., the $z$-axis magnetization profile itself is correlated in the $x$ and $y$ directions. In the language of XY model, the phase angle is also correlated under translations in the $x$ or $y$ directions.

Why then, one may wonder, do the series expansion results not converge to the right $\gamma$ value at high $\kappa$? Examining the limit $\kappa\rightarrow\infty$ suggests an answer. In that limit the next-nearest neighbor interaction dominates and the spins decouple into series of intercalated 1D Ising antiferromagnetic chains. That singular limit has 1D Ising universality for which $\gamma=1$. The finiteness of the HT series expansion thus probably results in a slow decay of $\gamma$ toward unity, as the large $\kappa$ terms in the series dominate the expansion. In this respect, the series is both a high temperature and low $\kappa$ expansion, which further restricts its range of validity.

%

\subsection{ANNNI Roughening Transition}
\label{sect:rough}
We first consider the roughening transition of the ANNNI model in the ferromagnetic regime (Fig.~\ref{fig:rough_sample}). Though the $T_R$ values extracted from simulations are quantitatively different from the series expansion results~\cite{berera:1993}, similar trends are observed (Figs.~\ref{fig:TKfreeenergy}). In particular, the transition temperature $T_R$ is relatively invariant to increases in frustration. The formation of an interface is not further stabilized by frustration, but rather decreases with increasing $\kappa$. 
And contrary to the scenario predicted for other microphase-forming systems, the roughening transition does not pass through or near the Lifshitz point~\cite{kahng:1990}. Instead, the roughening transition line on the
$T$-$\kappa$ phase diagram is expected to reach the FM phase boundary near $\kappa\simeq0.43$. Interestingly, a finite-temperature intercept suggests that
the FM transition may be notably different above and below $T_R$.

\begin{figure}
\includegraphics[width=\columnwidth]{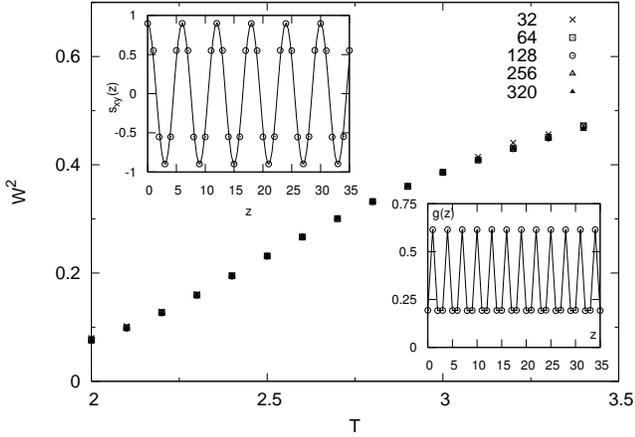}
\caption{Interface width $W^2$ of phase $\langle3\rangle$ of the ANNNI model at $\kappa=0.52$
 as a function of $T$ for various system sizes $L^2\times36$. 
 No roughening transition is detected within the stability regime. The magnetization profile $s_{xy}(z)$
 for the system of $L=128$ at $T=3.0$ is also shown (left inset) along with its normalized
 gradient $g(z)$ (right inset).}
\label{fig:rough}
\end{figure}

It has also been suggested that a roughening transition might be observed for the modulated phases as well~\cite{kahng:1990,levin:2007}. For the ANNNI model, however, we find no indication of interfacial roughening, at least for two simple modulated phases: phase $\langle2\rangle$ at $\kappa=0.8$ and phase $\langle3\rangle$ at $\kappa = 0.52$ (Fig.~\ref{fig:rough}). For the latter, in spite of reaching $T_c$, the interface location remains clearly defined with increasing system size. For any given temperature up to the PM transition, the interfacial width of the layers remains constant upon increasing the system size, but it is possible that a divergence can only be observed for much larger interfacial areas than what we consider. Yet the lattice is here at least an order of magnitude larger than the size necessary for detecting roughening in the ferromagnetic phase (Fig.~\ref{fig:rough_sample}).
We venture to speculate that, at least on a lattice, the persistence length of the lamellae might thus be very large and possibly infinite. If that were the case, the roughening of the modulated layers would then coincide with the PM transition. Further simulation and theoretical work are necessary to clarify the situation. 


\subsection{Phase Diagrams}
\begin{figure*}
\begin{center}
\includegraphics[angle=270,width=5in]{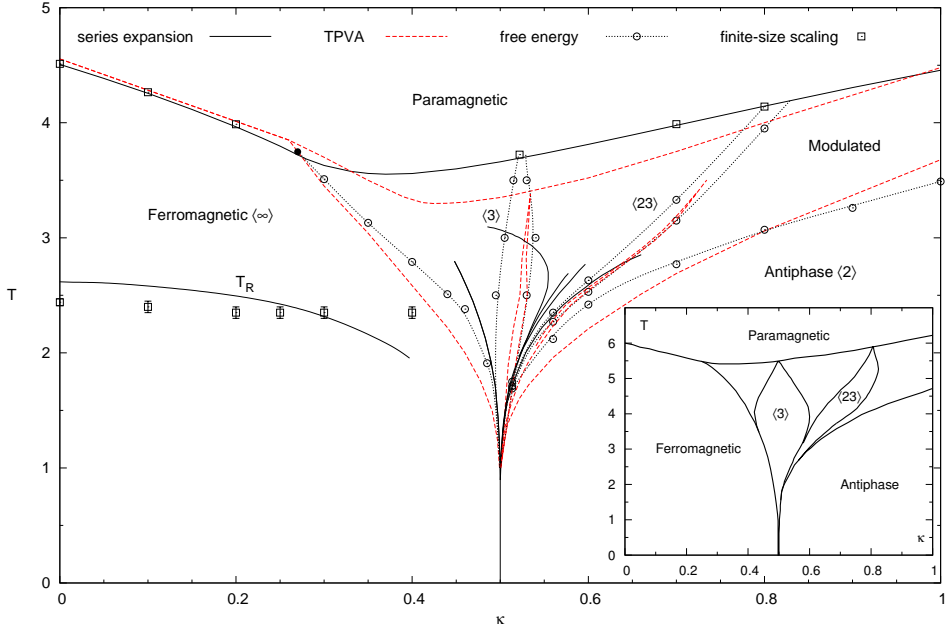}
\caption{Phase boundaries for the ANNNI model obtained from $U_4^*$ ($\Box$) and $F$ ($\odot$). The Lifshitz
point location ($\bullet$) is taken from Ref.~\onlinecite{pleimling:2001}.
High-~\cite{stanley:1977a,oitmaa:1985} and
low-temperature~\cite{fisher:1981} series expansions as well as TPVA~\cite{gendiar:2005}
results are represented. The stability wedges of phases $\langle 3\rangle$ and $\langle 23\rangle$ obtained from simulation are seen to be qualitatively different from the mean-field theory predictions (right inset)~\cite{selke:1984}. The roughening transition results in the ferromatnetic regime are similar to the series expansion results~\cite{berera:1993}.}
\label{fig:TKfreeenergy}
\end{center}
\end{figure*}

Detailed low $T$ series expansion studies of the phase behavior around $\kappa=1/2$ conducted by
Fisher \emph{et. al.}~\cite{fisher:1980,fisher:1987} suggest that a series of ``simple phases''
 of the form $\langle 2^j3\rangle$ spring out from the the multiphase point at $T=0$ and ``mixed phases'' generated by
 combinations of neighboring simple phases branch out at $T>0$. The temperatures accessible in simulations are relatively far from the regime of validity of this theory and thus, from this point of view, it is misleading to compare them directly. It
 is nonetheless interesting to note that the two approaches appear to converge for $T\lesssim 2$. 

Various approximate theoretical treatments have been used to analyze the ANNNI phase diagram more globally. In addition to the traditional mean-field approach~\cite{bak:1980}, an effective field~\cite{surda:2004} and a tensor product variational approach (TPVA)~\cite{gendiar:2005}
have more recently been used. These last two approaches reasonably capture the external boundaries
of the modulated regime.
The two treatments, however, qualitatively disagree on the internal structure of that same regime. On the one hand, the effective-field method~\cite{surda:2004}, like the mean-field treatment and the soliton approximation~\cite{bak:1980}, fills the modulated
interior by exceptionally stable bulging simple phases, 
such as
phase $\langle 3\rangle$ phase and phase $\langle23\rangle$ (Fig.~\ref{fig:TKfreeenergy}). On the other hand, TPVA predicts rather narrow stability wedges
for the commensurate phases~\cite{gendiar:2005}. The simulation results tend to favor the second scenario. Though the
devil's staircase indicates that the rate of wave number change slows on approaching $T_c$, only the antiphase has a broad presence in the modulated
regime. The stability range of the different modulations is fairly small, and all of the phases commensurate with
the periodic box are stable in turn. The branching and mixing of the stable low-temperature
$\langle 2^j3\rangle$ phases is already complete at the temperatures studied
here~\cite{fisher:1987,selke:1988}. In particular, no special stability is observed for phases
$\langle 3\rangle$ and $\langle 23\rangle$  (Fig.~\ref{fig:TKfreeenergy}). For phase $\langle 3\rangle$ some bulging is seen, because of the slower rate of change of the periodicity near $\kappa=1/2$. Simulating larger lattices, which allow for a more refined $q$ selection, however shrinks that phase's footprint. For phase $\langle 23\rangle$, the range of stability
does increase slightly with $\kappa$, but the effect is probably due to the finiteness of the lattice. In any case, the
increase is much less pronounced than the bulging scenarios predict~\cite{bak:1980,surda:2004} (Fig.~\ref{fig:TKfreeenergy}).

For the IC model, qualitatively similar branching is expected in the devil's flower region, which is found
between phases of integer periodicity. The systems simulated here are, however, too small to examine this
issue critically. Except for the caveats presented in the previous sections, our results mostly agree with the simulation results of
Ref.~\onlinecite{tarjus:2001}.


\section{Conclusion}
Our simulation study has clarified the structure and transition properties of the modulated regime of the ANNNI and the IC models. Previous theoretical treatments had sometimes been insufficient, particularly concerning the stability regime of the various modulated phases, the critical nature of the ANNNI PM transition, and the role of roughening. In the last case, no clear conclusion can be drawn, but the results suggest that the phenomenon is at least a lot less pronounced than in the Ising model, which may give hope of experimentally forming microphase patterns on much larger scales than previously thought. From a theoretical perspective, it is also interesting to highlight, however, that mean-field theory is particularly adept at predicting the periodicity of modulated phases at the PM transition. This observation may explain why the approach has been so successful at describing order in other microphase-forming systems, such as diblock copolymers~\cite{bates:1990,bates:1999}.


In addition to lamellar phases, modulated assemblies can exhibit a variety of other symmetries. They can also be observed off lattice. Generalizing the approach to continuous space and to other order types would thus greatly benefit the study of more complex microphase-forming systems. For the IC model, it could for instance help determine the nature of
the modulated-N\'eel transition and other properties of the high $Q$ regime, which we only briefly explored. Completing the simulation tool set would also pave the way for studies of the non-equilibrium microphase assembly, where most of the materials challenges lie.


\begin{acknowledgments}
We thank B. Mladek, J. Oitmaa, and M. Pleimling for their help at various stages of this project. We acknowledge ORAU and Duke startup funding.
\end{acknowledgments}


\begin{thebibliography}{123}
\bibitem[{foo({\natexlab{a}})}]{footnote:1}
\bibinfo{note}{For the paramagnetic phase, where formally $B_0=0$, the
  integration is equivalently taken from finite $T$ to infinity.}
\end{thebibliography}

\end{document}